# Evaluating autonomous self-healing against persistent cyber-attacks

Bart Gijsen[1], Stefano Nieuwenhuizen[2], Lucien Yap[2], Georgi Spasov[2], Erwin de Haan[1], Mahira Ali[1], Dr. Shari Finner[1]

[1] TNO Netherlands Organisation for Applied Scientific Research, Anna van Buerenplein 1, 2595 DA The Hague, The Netherlands
[2] Fontys Hogeschool, Rachelsmolen 1, 5612 MA Eindhoven, The Netherlands

**Abstract.** The need for autonomous cyber resilience (ACR) is strongly accelerating. The boost in cyber-capabilities of advanced AI is speeding up the attacker-defender rat race to an unprecedented pace. As a result, the field of ACR must progress from early design and experimentation to hands-on evaluation and deployment. To illustrate what is needed for this transition, we put TNO's autonomous immune system-inspired Self-Healing for Cyber Security (SH4CS) software to the test. Our analytic and experimental evaluation results demonstrate that effective cyber resilience is achieved with SH4CS for (repeated) web application vulnerability exploits. Remaining challenges of SH4CS revolve around finding the right balance between predictable resilient behaviour, and adaptiveness to the large variety of cyber threats and applications-to-be-protected.

## 1 Introduction

In the previous decade, the field of cybersecurity saw a steep increase in automated cyber-attack tools. To mitigate the increasing volume and speed of attacks in times of sustained scarcity of cyber expertise, also defenders put more effort into automating their cyber defences. Automated security technology, including SIEM-based threat and anomaly detection, automated vulnerability scanning, SOAR, DevSecOps and XDR tooling, was introduced and gradually adopted. In addition, more sophisticated, *autonomous* cybersecurity solutions have been introduced, such as fully automated patching, Run-time Application Self-Protection (RASP) and Artificial Immune System (AIS) technology. However, there was hesitation in the cybersecurity community to adopt autonomous security technology, in particular due to concerns that its deployment might disrupt the very digital services it is intended to protect.

While automation already led to an increased pace of the attacker-defender rat race, recently, the pace of this rat race has been boosted even further. We have seen the continuous build-up of (agentic) AI vulnerability detection and exploitation technology, where recent breakthroughs were made in traversing kill-chains [1]. Current AI-based exploitation progress has led to the point where fundamental cybersecurity practices can no longer keep up with the attacking side. Noteworthy is NIST's recent announcement: a 263% increase of CVE submissions between 2020 and 2025 forces NIST to change their way of listing CVEs in its National Vulnerability Database (NVD). From now on, only critical, known exploited vulnerabilities that have impact on software used by US government will be listed [2]. Moreover, it is becoming increasingly unrealistic to expect strongly reduced CVE patching times to turn the tide. More elaborate and creative defence-in-depth strategies will be required to keep pace in the accelerating rat race [3].

The need for deploying autonomous, defence-in-depth solutions has become urgent, and the cybersecurity community has started developing several (isolated) autonomous solutions. With the recent developments, the time has come to integrate solutions into an autonomous cyber resilience (ACR) architecture, evaluate implementations and make them ready for deployment. As a contribution to this task, in this paper we present an immune-system-inspired perspective on the ACR-architecture. Further, we present initial results of an evaluation of a partial ACR-implementation in the context of web applications. We conclude with a discussion about the next steps in our contribution to re-balance the attacker-defender rat race.

Although the development of ACR technology is spreading to other domains such as (autonomous) cyber physical systems, the focus of this paper is on the domain where most progress has been made to date: cloud-native environments.

# 2 Autonomous Cyber Resilience

## 2.1 ACR layers and related work

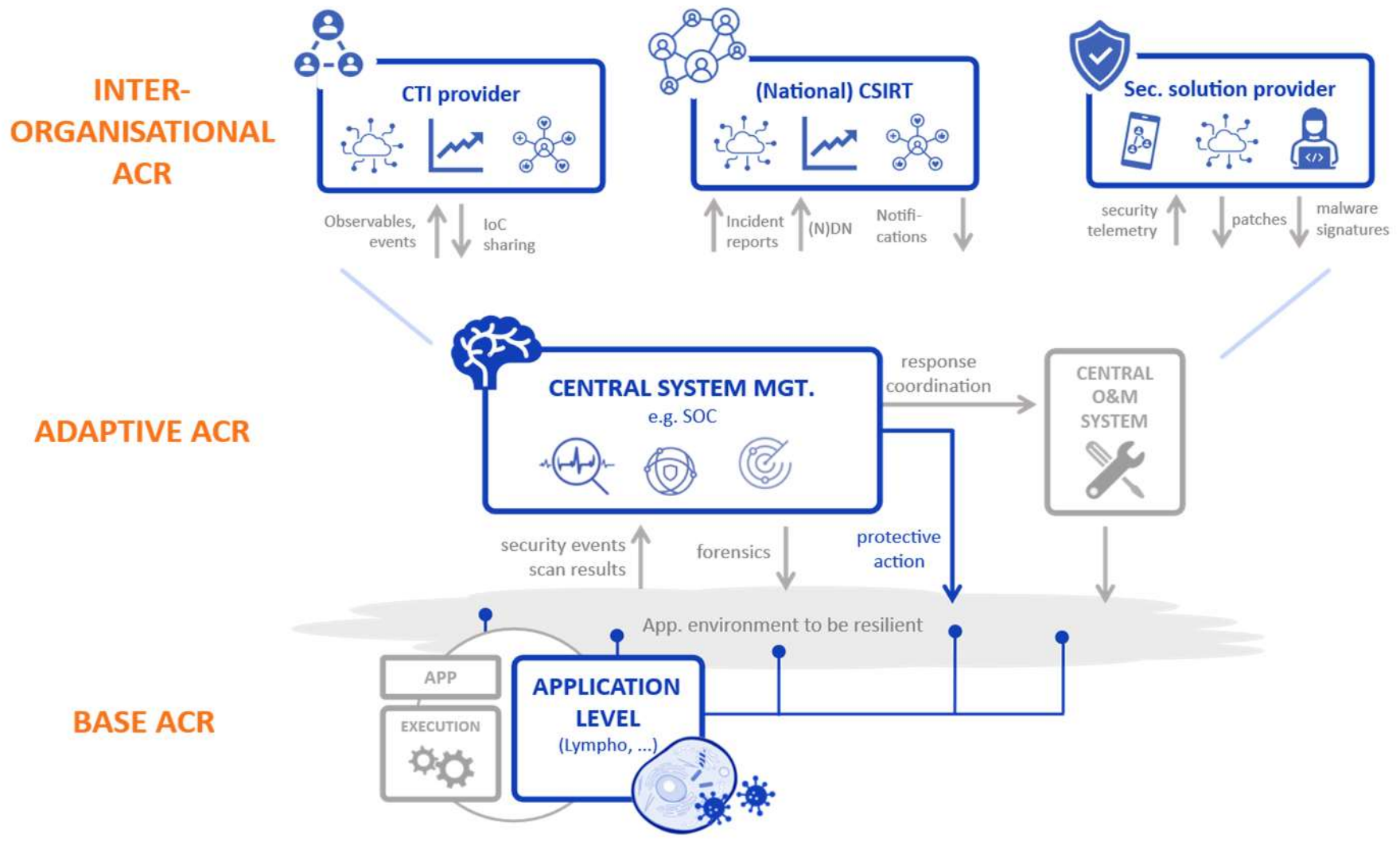

**Fig. 1:** Autonomous Cyber Resilience defence layers, inspired by the layers of the human immune system and human healthcare services.

Our perspective on ACR is based on the layers of defence of the human immune system, (the innate and the adaptive immune system), plus the surrounding societal ecosystem of healthcare, which includes interventions such as antibiotics, vaccination and other medical care. The analogy with the three biological defence layers allows us to structure and position our autonomous cyber resilience work as depicted in **Fig. 1** above.

The ACR techniques in each of the three layers are essentially control loops that perform repeated iterations of monitoring, analysing, planning and executing (MAPE). Similarly to the immune system, these control loops are nested and strongly interact with each other. More details about nested control loops and other fundamental ACR design principles can be found in [4].

**Base ACR - Self-Healing for Cyber Security (SH4CS).**
Initial development of Base ACR-technology started with the development of TNO's Self-Healing for Cyber Security (SH4CS) software [5] [6]. SH4CS extends the existing Kubernetes-native self-healing function, which periodically 'pings' each application-to-be-protected with liveness-probes and restarts its container when the application has become unresponsive and the probe fails. This simple mechanism can 'heal' an application that has become unavailable due to runtime application errors (e.g. memory leaks, data driven errors or buffer overflows). It is an example of RASP technology. TNO's SH4CS software extends the Kubernetes-native regeneration feature by placing a sidecar (called "*Lympho*" or "*Lymphocyte*") next to each application in the same pod. The Lympho captures Kubernetes' liveness-probes and runs configurable, match-action rules to determine whether or not Kubernetes will be returned a *healthy* or *failed* probe, thereby effectively controlling container regeneration. A container restart not only heals run-time errors but also wipes (potentially undetected) malware infections and attacker footholds that are stored in memory or in temporary files.

Next to capturing probes, the Lympho also keeps track of the application's health-state, which can be modified and updated according to its configuration rules. This way, the Lympho is a simplistic building block for integrating ACR mechanisms such as:

- security monitoring alerts received from cloud-native tools, e.g. Falco or Prometheus, which trigger match-action rules;
- web application firewalls (WAFs) that control all traffic to/from the application based on the Lymphocyte's health state;
- periodic container regeneration mechanisms, preventing several types of runtime errors and wiping undetected, maliciously injected footholds;
- automated patch deployment using DevSecOps tools to automatically spool in new, patched container images at times of Lympho-initiated container restarts.

In analogy to the immune system, the role of SH4CS as a "*Base ACR*" mechanism (see **Fig. 1**) can be regarded as an innate immune function that provides a generic, immediate response without analysing the root cause. As stated in [3]: "RASP and WAF are meaningful *interim* mitigations against new or even unpublished CVEs." In order to mitigate the root cause of CVEs, more intelligence is required. In the immune analogy, any thorough analysis and cyber-reasoning relate to the more time and energy consuming targeted response of the *adaptive* immune system, as described in the following section.

**Adaptive ACR.**

Cloud-native environments are traditionally managed and protected centrally from Security Operations Centres (SOCs). The used security tools and procedures are typically advanced, but often still *reactive* to detect, analyse and mitigate cyber threats and attacks. The evolution from the traditional SOC to an automated and autonomous SOC is described in [7], where the author points out that improvements in AI reasoning, real-time orchestration, and agent collaboration for SOC operations are still to be made.

Further improvements are required around autonomous cybersecurity technology that focuses on fully automated patching [8] and autonomous cyber forensics [9]. Forensics is a typical interface between *Base ACR* and *Adaptive ACR*, similar to the mechanism of *innate* immune antigen presentation mechanisms that are used to activate targeted responses of the *adaptive* immune system. Currently, we are further exploring autonomising such mechanisms to expand the SH4CS software suite, including the development of a "*cyber immune memory system*".

The nesting of *Base ACR* into *Adaptive ACR* control loops is organised via the following most relevant types of interfaces:

- *Adaptive ACR* may adapt (e.g. patch, reconfigure, reassemble) the Application-to-be-Resilient (A2bR), which is then deployed into operation by *Base ACR* techniques such as periodic or trigger-based container regeneration.
- Container image adaptation/updates by *Adaptive ACR* may be triggered by forensic data collected by *Base ACR* observability tools.
- *Adaptive ACR* may modify *Base ACR* rules and config parameters (see **Section 4.2**).

**Interorganisational ACR.**

In case a human's immune system falls short, or to improve the immune response of a population, society employs the medical care system. It encompasses medical knowledge about the immune system and interventions that can be taken, such as prescribing antibiotics, antiviral drugs or organising vaccination campaigns. Similarly, the concept of *interorganisational ACR* may be thought of as a means to further improve organisations' resilience mechanisms already in place. Examples are (national) CSIRTS exchanging cyber threat intelligence (CTI), security service providers performing red-teaming, cyber forensics, etc., or cybersecurity insurance companies applying cybersecurity related conditions.

This diverse ecosystem is also raising its service level autonomy: for instance, progress is being made in autonomous red teaming [10], while the challenge of autonomous decision-making for cyber resilience is still an active field of research [11]. Also, the concept of active cyber defence (ACD) is increasingly used and being promoted by national CSIRTs. In [12], an Autonomous Cyber Defence System Architecture is developed, which resembles the lower two layers of the ACR architecture of **Fig. 1**. Other authors propose governance frameworks favouring cyber defenders by creating asymmetric frontier AI-model access models [13]. Similarly to the medical care analogy, potential interventions in the interorganisational ACR layer are quite diverse.

### 2.2 Challenges

**The gain should outweigh the pain, under unpredictable circumstances.**
The purpose of autonomous cyber resilience is to create (digital computing) systems that can anticipate, withstand, recover from, and adapt to cyber-attacks and (digital) malfunctions to ensure mission or business objectives without being controlled directly by humans. ACR techniques may have a direct impact on mission or business objectives – any erroneous or abused ACR can have negative business or mission impact. The challenge is to ensure that under all circumstances – in particular adversarial circumstances – applying ACR results in a net positive improvement on business or mission objectives. How to specify and quantify these objectives and improvements well is yet another open research question.

**Rapidly building trust in autonomous cybersecurity.**
Five years ago, autonomous cybersecurity publications such as [5] were received with cautious restraint by the cybersecurity community. While AI-based security monitoring was gradually being adopted, autonomous mitigation that may disrupt digital services was met with strong hesitation. Building trust to fit active autonomous cyber tooling into common cybersecurity strategies takes time; yet time is also running out.

In addition, AI will play a significant role in autonomous cybersecurity. While vast progress is being made at this front, progressing insights also show that AI-assisted defence introduces its own challenges. In [12] these are summed up as: data availability and quality issues, false positives and negatives, model drift (e.g. due to rapidly changing attack patterns), adversarial attacks and limited explainability. For example, improvements are required for autonomous defence agents to make them both applicable under realistic cyber offensive circumstances [14] and robust to adversarial attacks [15].

## 3 SH4CS evaluation approach

### 3.1 Evaluation objective

The main question is: How can ACR technology be developed to be effective against state-of-the-art offensive AI capabilities? The effectiveness of SH4CS against specific, single-stage attacks on web applications has been proven in earlier work [5]. However, AI-assisted attacks focusing on vulnerability detection and exploitation of internet-facing APIs are no longer limited to single-stage attacks [1] [16]. Our endeavour to develop ACR technology that matches novel offensive AI capabilities starts by exploring how TNO's existing SH4CS software behaves against persistent multi-stage attacks.

### 3.2 Evaluation set-up

In this paper we take a repeated, multi-stage crypto-mining attack as a starting point for further research on increasingly complex AI-assisted attacks. As the Application-to-be-Resilient (A2bR) and the target of attack, we select the Damn Vulnerable Web

Application (DVWA) [17]. This application is intended for experimental use in the cybersecurity community. It is built from real-world software products: Linux, an Apache server, MariaDB and a PHP interpreter (enabled for the Apache server), but with slightly outdated software versions of those products that contain known vulnerabilities. We consider a setup with four load-balanced Apache container instances (each with a PHP interpreter as part of the container) and one MariaDB container deployed in a single pod on a Kubernetes cluster. Each of the four Apache containers is protected by a SH4CS Lymphocyte. This four-fold 'redundancy' is used to ensure that, during regeneration of one container, the remaining three containers are able to handle all incoming requests.

Each Lympho is configured with a single rule: once a sequence of attack steps is observed that includes (a) either an initial access technique or a credential access technique, and (b) a malicious code execution technique, then the liveness probe is not responded, triggering an application container regeneration. For detection of the attack techniques, the following cloud-native tools in the Kubernetes cluster are used:

- Prometheus, which periodically extracts Apache request log records and maps them to MITRE ATT&CK techniques;
- Alert Manager, which tracks the changes generated by the MITRE ATT&CK technique sightings in Prometheus, and groups and forwards alerts to the corresponding Lympho sidecar.

While in principle, it is also possible to include WAF functionality as part of the traffic load balancer, the focus of this evaluation is mainly on the Lympho sidecar's container regeneration feature.

### 3.3 Attack model

**Multi-stage attack.**

We consider an attacker who noticed a vulnerability in the Apache server software and tries to exploit containers for crypto mining. They use an agentic AI attack-tool that persistently tries to infect the containers with malware performing crypto mining. Once the malware starts crypto mining, it consumes all processing resources assigned to the Apache container, so that the container can no longer fulfil its intended functionality. From the defender's perspective, this crypto-mining attack is a denial-of-service attack.

To achieve their malicious intent, the attacker developed a tool containing several attack scripts and an LLM-agent. The first script exploits the identified vulnerability in the Apache application (MITRE ATT&CK technique T1190), delivering crafted requests that trigger remote command execution within a container. Upon successful exploitation, it leverages command and scripting interpreter capabilities (T1059) to download and execute a crypto-mining payload. The LLM-agent iteratively adjusts its attack scripts based on observed responses, increasing the likelihood of successful execution across varying container states. Once deployed, the miner starts consuming system resources and making mining results available for download by the LLM agent. Besides mining, the deployed payload is also used for scanning the environment in which the infected container is deployed. In our DVWA setup, three other containers running in the same pod may be discovered. When the payload reports this back to the LLM-agent, the agent tries to further spread its payload infection to all other Apache containers from within the pod.

**Repeated attacks.**
A generic challenge for developing autonomous cybersecurity is the limited detailed attack data available. Generic attack patterns are published, such as MITRE ATT&CK, but those lack details about, e.g., attack step timing and repetition of steps in case previously established footholds are wiped. Moreover, such details are typically very attack-case-specific. However, looking at [16], we assume that it is (or will become) likely that AI-assisted attack tools are 'intelligent' enough to verify the success of their attack, and will aim to repeat any steps nullified by a SH4CS self-healing action.

In the context of the multi-stage attack mentioned above, the LLM-agent periodically downloads crypto-mining results and verifies whether crypto miners are still running. If not (as container regeneration by SH4CS has wiped the payload), the agent reinitiates the T1190 + T1059 exploitation cycle. The attack interval period configured for subsequent LLM-agent invocations depends on the attacker's preferred balance between securing crypto-mining results and the risk of being detected and wiped.

### 3.4 SH4CS optimisation

Given a specific attacker model, a self-healing optimisation problem emerges: If a container gets infected, the SH4CS logic actively drops Kubernetes liveness-probes, initiating a stop of the infected container and restarting of a clean container image. Each container will be regenerated at the latest upon its lifetime expiry, or any time sooner if an infection is detected. Note that in the context of the crypto-mining attack, the default container lifetime is much larger than the attack rate, such that in most cases a timer expiry does not occur before an infection is detected. Kubernetes liveness-probes are generated with fixed time-intervals of 15 seconds by default. This, amongst other aspects, controls the workload overhead for running the Lympho sidecars executing the SH4CS logic. The longer the configured probe interval, the smaller the regeneration overhead, but the longer the average infection detection time. In combination with the timing behaviour of the attacker, the following optimisation question arises: which probe-interval time maximises the average number of active, uninfected containers?

**Quantitative attack-defence problem.**
A common approach for modelling targets that are subject to 'self-evolving' malware infections is the SIRS model [18], borrowed from the field of epidemiology. It models a 'target' as being in one of three states: Susceptible, Infected or Recovered, and possible state transitions between these three. Thus, SIRS models the attacker and defender actions in an integrated (composed) process. Further, it models targets independently from each other. To analyse the impact of the SH4CS parameters, an extension of the SIRS model is required where attacker and defender processes are decoupled, and the SH4CS parameters are included in the state transitions. Separating the attacker and defender also provides the opportunity to distinguish between initial container infection by the attack tool and subsequent spreading of infections within a pod from one container to another.

This attacker and self-healing optimisation problem can be modelled using stochastic process algebra (SPA) [19]. SPA is a method for analysing the composed behaviour of interacting 'processes'. In the crypto-mining case, these processes include the attacker(s) and the application containers, each of which is protected by a SH4CS Lympho software instance.

**Fig. 2** presents the adapted SIRS state transition diagrams that enable analysis of the liveness-probe interval (denoted as 'SH4CS scanning interval'), provided the attack parameters 'Time to compromise container' and 'Time to spread malicious payload to other container' are given. To investigate this quantitative parameter optimisation problem, we applied a combination of experimentation and Markov reward optimisation.

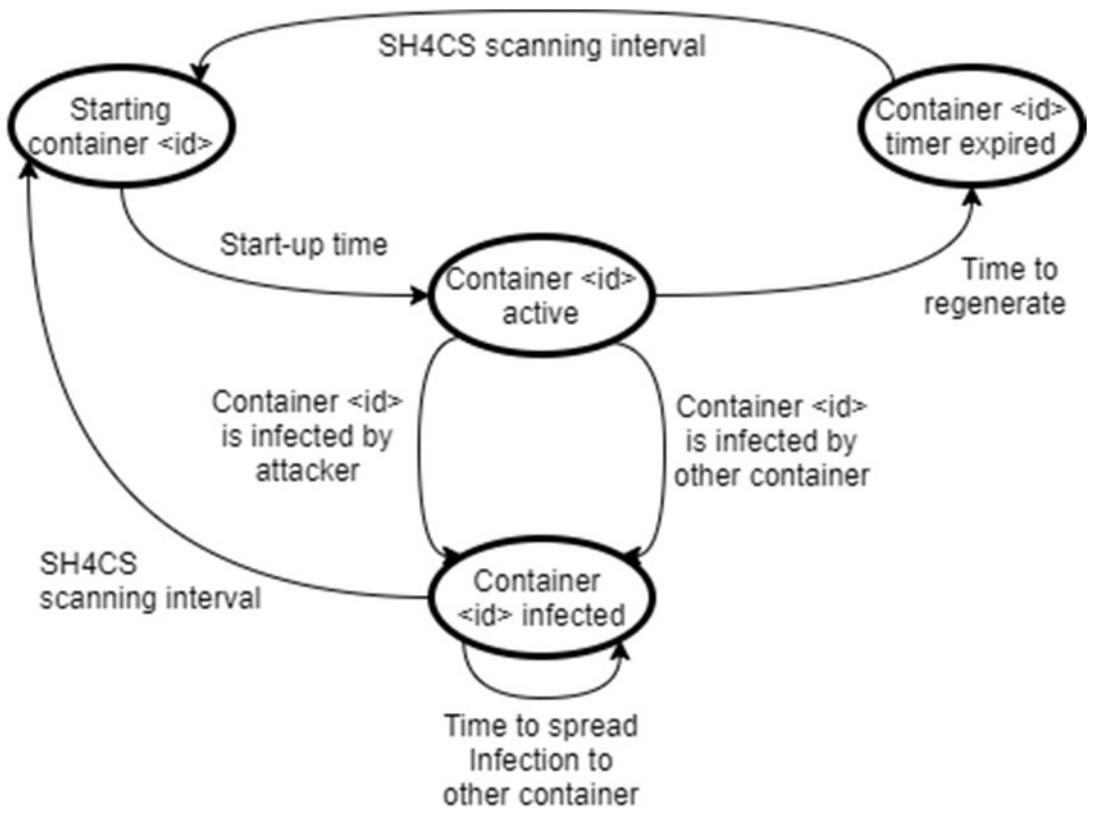

**Fig. 2:** Combined attack and SH4CS state transition diagram (single container)

# 4 Results

## 4.1 Hacking DVWA and combatting SH4CS

To experimentally validate the crypto-mining case with SH4CS protection, we used the experimental set-up to verify (a) the feasibility of the attack steps, and (b) the behaviour of an automated attack tool that repeatedly attacks the SH4CS-protected DVWA application. During these experiments we also measured parameters such as a container regeneration time (i.e. the start-up time in **Fig. 2**).

Our experiments only yield initial, preliminary results of our investigation of the effectiveness of SH4CS against persistent attacks. Limitations of our experimental setup are, for instance, the fact that we did not implement four Apache server replicas, and that the number of repeated attack iterations is limited (see discussion in **Section 5**). Furthermore, our attack scripts were written manually, the LLM-agent of the attack tool was not (yet) used to produce the results presented in this section. An LLM-agent only becomes relevant for more sophisticated experiments, crafting e.g. pod-specific cross-container infections. This will be performed in later phase of our investigation.

**Feasibility of multi-stage crypto-mining attack.**

In the documentation of the (intentionally vulnerable) DVWA application, examples are provided on how to perform command injection attacks (T1059) in the shell of the Apache container [17]. Exploiting this vulnerability, the attacker can abuse an unprivileged shell (www-data) to discover open ports, accessible endpoints and exposed services on the shared loopback interface of the pod. Even without prior knowledge, implementation details of other containers and of the underlying Prometheus and Kubernetes infrastructure can be observed. Sufficient details can be extracted to craft new attack PHP scripts to perform further monitoring and crypto mining, and to attempt break-out to infect other containers in the pod. While PHP scripts were developed manually in our experiment, such scripting is well within the capabilities of modern LLMs.

In conclusion, manual or autonomous execution of the multi-stage crypto-mining attack is certainly feasible, and we observed substantial attacker manoeuvrability resulting from an exploitable command injection vulnerability, a shared namespace (in a pod) and basic LLM capabilities. We also learned that the Lympho sidecar, Prometheus interface and Alert Manager will require their own authentication and authorisation configuration to avoid exploitation of the SH4CS mechanism itself by the attacker.

**Behaviour of repeated attack and defence.**
Our second type of experiment focused on the behaviour of the attack tool and SH4CS regenerations over consecutive attack waves. Experiments with two different two-stage attacks were performed. In scenario 1, an initial access technique is followed by the deployment of an SQL injection. In scenario 2, the initial access technique is followed by a command injection (T1059), i.e., the malicious code deployment technique of the crypto-mining case. In both scenarios the initial access technique chosen was brute force (T1110), which is different from the initial access technique T1190 used in the crypto-mining case, but results in the same effect. For this second type of experiment, no crypto-mining script was uploaded, since it added little to the insight to be gained.

After each two-stage attack, a period follows in which SH4CS detects both attack stages and regenerates the Apache container. Each run was measured for two consecutive, autonomous attack-defence cycles. For both scenarios, ten runs were executed to verify the consistency of the results.

The measurements in all runs indicate that each attack step was successful, each of them was detected and the SH4CS-triggered regeneration wipes any deployed malicious code. In **Table 1,** the relevant time measurements are shown. The presented statistics are the detection and regeneration times for the first attack step, since these provide the appropriate empirical input for the crypto-mining case[1].

**Table 1.** Event timing (average +/- standard deviation across 10 runs).

| KPI | 10 Runs of scenario 1 | 10 Runs of scenario 2 |
|---|---|---|
| Detection time | 3.2 +/- 1.1 s | 3.8 +/- 0.9 s |
| Regeneration time | 9.1 +/- 1.7 s | 9.6 +/- 1.8 s |

[1] For the repeated attack step, the measurement of the detection time is ‘blurred’ by specific suppression parameters in the SH4CS Lympho sidecar, as explained in **Section 5**. Averaging the detection times over the first and the repeated attack steps would result in inappropriate empirical input for the crypto-mining case.

The detection of attack steps is not instantaneous for two reasons: First, the multi-step attack itself is not instantaneous. Although automated, two subsequent attack steps need to be successfully executed after each other. This leads to some fractional delay. Second, as explained in **Section 3.2**, the detection of each attack-step is not instantaneous. Both observability tools, Prometheus and Alert Manager, are configured with data collection intervals. In our experiments Prometheus is configured with a 1 second interval to scrape the Apache log file for new events. In addition, the Alert Manager is configured with a 5 second group interval. The sum of those two intervals leads to a theoretical average detection delay of around (1+5)/2 = 3 seconds. These two additional types of delay explain the detection results presented in **Table 1**. The measured regeneration time of nearly 10 seconds is typical for container regeneration in cloud-native environments and has been empirically determined in earlier publications [5].

### 4.2 Markov reward optimisation for SH4CS

**Quantifying model parameters.**
As indicated in **Section 4.1,** our experimental results do not yet cover the full scope to answer the optimisation question which probe-interval time maximises the average number of active, uninfected containers. Nevertheless, they provide empirical input to quantify several parameters of the SPA model. Our intention is to validate the model with more elaborate experimental results in a later investigation.

Particularly, the 'SH4CS scanning interval' (see **Fig. 2**) is the liveness probe-interval[2], i.e., the sum of the detection time and the optimisation parameter. Vice versa, the empirically determined detection time (see **Table 1**.) should be subtracted from the optimised value for the SH4CS scanning interval to obtain the to-be-configured liveness probe-interval maximising the mean number of active, uninfected containers. Moreover, the 'start-up time' (see **Fig. 2**) equals the empirically determined regeneration time in **Table 1**. In the SPA model, a start-up time parameter value of 10s is used.

For the application of the SPA model to the crypto-mining case, several attack pace scenarios are analysed, which are distinct regarding the following aspects:

- whether the attack tool succeeds in **compromising** a container fast (within 1 minute), medium speed (10 minutes) or slow (within 1 hour) and
- whether the payload **spreading** from one infected container to another is fast (within 15 seconds) or medium (within 5 minutes).

The timing parameter values are selected in such a way that their combination covers varying crypto-mining attack strategies in which the attacker balances the priorities of obtaining successful results quickly vs. reducing the risk of detection. After analysing all possible combinations of infection and spreading rates, the following four attack scenarios appear to be most 'aggressive': (1) fast compromise and fast spreading, (2) medium compromise and fast spreading, (3) fast compromise and medium spreading and (4) slow compromise and fast spreading.

[2] Note that the SPA model is simplified by using only <u>one</u> SH4CS scanning interval. In particular, the detection time should not be included in **Fig. 2** in the transition from "container timer expired" (detected instantaneously) to "starting container". However, in the persistent attack crypto-mining case, the state-diagram hardly ever reached the state where a container regeneration is triggered by a timer expiry. In other, less aggressive cases, the detection time is typically small relative to the liveness probe interval. Ergo, this simplification does not significantly affect our results from the SPA model analysis.

**Optimising the reward.**
Given the state transition diagrams of the attacker and the stochastic behaviour of the four SH4CS-protected Apache replicas, the composed processes can be modelled as a Continuous Time Markov Chain (CTMC), under the assumption that the time intervals are exponentially distributed. We implemented the state diagrams in the SPA tool "PRISM" [20], which automatically generates the CTMC. CTMC generation requires filling in a specific value for the SH4CS scanning interval parameter, i.e., the parameter to be optimised in our study. To estimate the optimum, we selected a range of values for the scanning interval parameter between 1 and 30 seconds, and generated CTMCs for each parameter value in this range. Subsequently, the PRISM tool calculates the steady-state probabilities of all composed states for each CTMC. From the steady-state probabilities, the time-average number of active uninfected containers (the reward function) can be calculated. Our goal is to have the reward function value remain ≥ 75%, such that on average at least three out of four Apache containers are active and uninfected to handle the computational load.

**Optimisation results.**

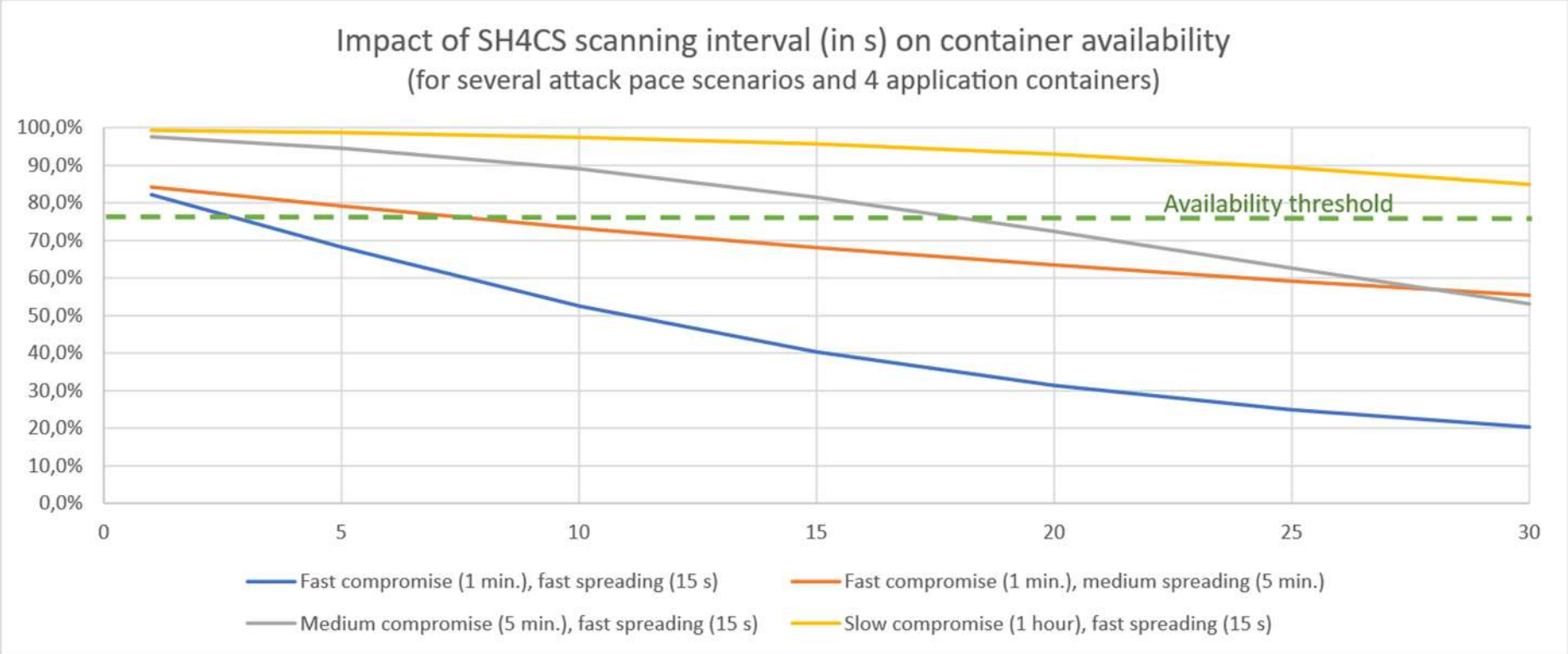

**Fig. 3.** Impact of the SH4CS scanning interval on container availability.

The calculated container availability with the SPA tool for each scenario is presented in **Fig. 3**. It shows that, in all scenarios, a SH4CS scanning interval is found that is at least as large as the average detection time; i.e. in all investigated scenarios a 75% container availability requirement is feasible. The most aggressive attack scenario (fast infection and fast spreading) is the corner case in which the SH4CS scanning intervals should be configured at a value as small as 3 seconds (which is in the same order as the detection time) to ensure container availability of at least 75%. This implies that, although the SH4CS computation overhead for collecting monitoring data and determining whether regeneration is required is relatively small, it may become significant when coping with such aggressive attack scenarios. In contrast, for the attack scenario of medium time to compromise and fast spreading, a SH4CS scanning interval of 18 seconds is sufficient, at which SH4CS computation overhead becomes negligible.

While this Markov reward model analysis is based on static parameter values, it illustrates that dynamic parameter optimisation which adapts the parameter to the pace of attack at any specific point in time, is sensible. Typically, an *Adaptive ACR* functionality to be implemented would automatically assess the attack pace and re-configure the SH4CS parameter to an optimised value.

# 5 Discussion

## 5.1 Challenge 1: The gain should outweigh the pain

The results in **Section 4.2** show that SH4CS, in combination with replication redundancy, provides effective autonomous cyber resilience for the DVWA against the multi-step, repeated crypto-mining attack. Although the investigated adversarial behaviour is rather specific, it demonstrates that a simple rule-based integration of the resilience measures of container replication, cloud-native security monitoring and container regeneration can provide a controlled, autonomous response. Moreover, the rule complexity in this crypto-mining case is linear in the number of different attack techniques and the chosen course of action (i.e. container regeneration). While this first exploration is barely scratching the surface of autonomous resilience mechanisms against automated vulnerability exploitation of internet-facing applications, it is a promising start.

**The suppression dilemma**.
Next to the promising results, our experiments showed some interesting, unforeseen practical issues. Here, we briefly address two anecdotal issues that may provide a glance into a fundamental dilemma of autonomous (cyber and immune) resilience:

The Lympho rule engine is stateful and tracks multiple temporal behaviours. When protecting against parallel, multi-stage attacks, it is by-design required to track some (abstraction of) attack progression. To avoid an unbounded and uncontrolled escalation of the threat level, rules are required to lower the Lympho's threat-level after some time – the immune system's equivalent would be the 'immune suppression' mechanism. In our current experiments, we used the open source SH4CS version 2.0 (2021) with a very simple suppression mechanism: once an attack technique is detected, the threat level is increased, and a timer is set to decrease the threat level after a configurable time window (in the experiments: 60 seconds). Within this time interval, any other detected occurrence of the same attack technique is ignored. In our crypto-mining case, this simplified suppression mechanism suppresses the detection of a second identical attack-step within 60s of a previous attack. A simple remedy for this issue would be to decrease the configurable 'threat increment duration' parameter value, or to use more sophisticated suppression methods such as leaky buckets. More sophisticated methods may yield better detection results. However, they may also increase ‘anxiety’ (overreaction of the immune mechanism) and/or significantly add to the system complexity and overhead. An alternative solution may be to introduce an (RL-based) AI agent instead of the rule-based Lymphocyte. However, looking at results such as those presented in [15], we are sceptical that this will resolve the immune-suppression dilemma in an environment with temporally varying threat levels. In multiple debates among cyber resilience experts, a vast number of cases have been identified that each would require a slightly different and tailored immune suppression approach.

The second ‘practical issue’ arose when varying the number of repeated attacks to investigate whether and how SH4CS resilience effectiveness might degrade over multiple cycles. This unexpectedly appeared to be the case due to a native feature of the Kubernetes self-healing function: Kubernetes keeps track of how often a specific container is being restarted in a certain time period. If the restart rate exceeds a threshold, Kubernetes itself starts to actively delay container restarts; a feature known as *CrashLoop-Backoff*. The Lymphocyte in our crypto-mining experiment appeared to be limited to perform exactly *five* repeated regenerations, after which the regeneration got delayed due to the CrashLoopBackoff. The abovementioned immune suppression dilemma thus also holds for the simple Kubernetes-native self-healing function. Related investigations of self-healing in a commercial PLC platform[3] observed the exact same practical issue in regeneration behaviour, demonstrating that self-healing effectiveness and immune suppression are inextricably linked phenomena.

It remains a fundamental open research question and cyber-resilience challenge how to address SH4CS suppression. In the human immune system, the adaptive immune suppression mechanism strongly regulates the effectivity of the immune system, striking a very fragile balance between immune system overreactions (e.g. allergies or other immunity disorders) and underreactions (e.g. immunosuppression under stress, allowing infections to thrive). Noting that millions of years of biological evolution have not been able to develop a universal solution resolving the immune suppression balance for all cases, we expect that also in our cyber resilience expertise field, developing a ‘silver bullet’ to resolve this challenge universally for all applications and attack scenarios may not be feasible.

**The need for inter-regeneration adaptiveness.**
In this paper we present initial results of the evaluation of SH4CS (as part of *Base ACR*) against the multi-step, repeated crypto-mining attack. While effective, it is not efficient. In the current setup, the SH4CS vs. crypto-miner battle would run indefinitely. SH4CS is adaptive within a regeneration cycle, but lacks adaptiveness and improvements by learning *between* regeneration cycles. To ‘win’ the battle, ACR needs a way to take away the starting points for the attack, which in our example is the command injection vulnerability in the DVWA. Here is where nested control loops between different ACR levels come in. A complementary, *Adaptive ACR* technique would be required to, e.g., patch the DVWA and/or adjust its firewall rules to tighten request injection validation. We are currently exploring autonomous forensic technology that is capable of (a) capturing relevant, 'pre-mortem' container logs, syscalls, configuration settings and adversarial metrics, (b) processing those and mapping them to MITRE ATT&CK techniques, (c) proposing appropriate mitigations for improvements of the container image. The large challenge and open research question in this chain revolves around evaluating several potential mitigations to select the most effective one.

[3] PLCs (Programmable Logic Controllers) are systems that control cyber-physical systems such as robot arms, conveyor belts, bridge hydraulics, etc.

### 5.2 Challenge 2: Rapidly building trust

In this paper, we mainly investigated the performance of the technical SH4CS solution and did not explicitly address the building of trust in ACR tooling that may disrupt parts of the system functionality. Our experience shows that developing trust requires the human to be in control at all times, i.e., the supervisor responsible for the environment needs to be able to instantly (a) disable potentially service disruptive ACR measures (e.g. set ACR to 'monitor and log only') and (b) to disable or undo ACR adaptations (a change-freeze period or roll-back) and to reset *Base ACR* to a validated and approved configuration profile. Rule (a) is a measure to control the autonomy level of ACR, while rule (b) should mitigate any additional risks introduced by AI-based decision making in ACR, provided that the AI-tooling is primarily used in the *Adaptive ACR* layer of **Fig. 1** and avoided in the *Base ACR* layer.

## 6 Conclusion

In this paper, we have explored the role of ACR in addressing the accelerating attacker–defender dynamics driven by advances in AI-enabled offensive capabilities. Using the SH4CS software as a concrete implementation of *Base ACR*, we evaluated its effectiveness against persistent, multi-stage cyber-attacks in a cloud-native web application environment. The experimental results demonstrate that SH4CS, in combination with replication and monitoring, is capable of consistently detecting and mitigating repeated attack cycles through automated container regeneration. Moreover, a quantitative analysis with stochastic modelling provides insight into the trade-offs between detection latency, regeneration overhead, and system availability under varying attack scenarios.

Our findings show that even relatively simple rule-based self-healing mechanisms can yield substantial resilience against repeated exploitation attempts, given that configuration parameters are tuned to the pace of the attack. Yet, the results highlight that static configurations are insufficient in the face of dynamic and adaptive adversaries, motivating the need for adaptive parameter optimisation. Furthermore, the suppression dilemma underlines that effective ACR is inherently linked to balancing priorities between responsiveness and stability.

SH4CS in its current form provides a strong reactive resilience capability but does not eliminate the root causes of vulnerabilities. This limitation emphasises the necessity of integrating *Base ACR* with higher-layer *Adaptive ACR* mechanisms, such as autonomous forensics and patching, to achieve sustainable resilience. In this context, the analogy with the human immune system remains a useful guiding principle where immediate responses are complemented by longer-term adaptive improvements.

Overall, this work illustrates that autonomous self-healing can play a pivotal role in rebalancing the attacker–defender race, but also that its effectiveness depends on careful orchestration across multiple layers of ACR. Continued research is required to improve inter-cycle adaptiveness, enhance trust in autonomous decisions, and integrate self-healing with broader cyber defence ecosystems.

**Acknowledgments.** This study was funded by TNO’s Exploratory Research Program “Cyber Secure Systems by Design” and performed in collaboration with Fontys ICT Eindhoven.

**Disclosure of Interests.** The authors have no competing interests to declare that are relevant to the content of this article.